\documentclass[aps,prb,floatfix,epsfig,twocolumn,showpacs,preprintnumbers]{revtex4}
\usepackage{amsfonts}
\usepackage{amssymb}
\usepackage{graphicx}
\usepackage{amsmath}

\begin{document}

\title{Full self-consistency versus quasiparticle self-consistency in diagrammatic approaches: Exactly solvable two-site Hubbard Model}
\author{A.L. Kutepov}

\affiliation{Ames Laboratory USDOE, Ames, IA 50011}

\begin{abstract}
Self-consistent solutions of Hedin's diagrammatic theory equations
(HE) for the two-site Hubbard Model (HM) have been studied. They
have been found for three-point vertices of increasing complexity
($\Gamma=1$ (GW approximation), $\Gamma_{1}$ from the first order
perturbation theory, and exact vertex $\Gamma_{E}$). The comparison
is being made when an additional quasiparticle (QP) approximation
for the Green function is applied during the self-consistent
iterative solving of HE and when QP approximation is not applied.
The results obtained with the exact vertex are directly related to
the presently open question - which approximation is more
advantageous for future implementations - GW+DMFT or QPGW+DMFT. It
is shown that in the regime of strong correlations only originally
proposed GW+DMFT scheme is able to provide reliable results. Vertex
corrections based on Perturbation Theory (PT) systematically improve
the GW results when full self-consistency is applied. The
application of the QP self-consistency combined with PT vertex
corrections shows similar problems to the case when the exact vertex
is applied combined with QP sc. The analysis of Ward Identity
violation is performed for all studied in this work approximations
and its relation to the general accuracy of the schemes used is
provided.

\end{abstract}

 \maketitle

\section{Introduction}
\label{intro}

One of the challenges for the computational theorists working in the
solid state electronic structure field is the robust implementation
of the so called GW+DMFT method (combination of GW approximation (G-
Green's function, W - screened interaction) and Dynamical Mean-Field
Theory). The scheme was originally proposed by Sun and
Kotliar\cite{prb_66_085120} and in slightly different (but probably
more commonly known) form by Biermann et al.\cite{prl_90_086402} The
basic idea of the approach is to separate all active space of the
basis set into "weakly correlated" part for which GW approximation
is supposed to work well and "strongly correlated" part for which
one sums up many diagrams (to infinite order if one uses the
non-perturbative DMFT solver as for example the Quantum Monte Carlo
- QMC). In order for the method to be well defined, everything
should be done till full self-consistency (sc), including the
iterations of the GW part itself and also the "internal" iterations
of DMFT part to ensure that the solution of the impurity problem
reproduces the same G and W in strongly correlated subspace as the
ones in the same subspace projected from the GW part.

Despite its obvious appeal GW+DMFT has made only slow progress
during more than decade since its first appearance in the above
mentioned papers.  The reason was not just because scGW is quite
demanding computationally but mostly because one has to satisfy the
impurity sc condition not only for G (as in the LDA+DMFT method - a
combination of Local Density Approximation in Density Functional
Theory and DMFT) but also for the W, which seems to be not an easy
task. To the best of my knowledge there were only "one-shot" type
calculations for real materials\cite{prb_90_165138} where GW
iterations were neglected altogether and DMFT self-consistency was
imposed only on G, whereas W was fixed at LDA level and
correspondingly the U was considered as an external parameter
(calculated in constrained random phase approximation - cRPA). Such
an implementation, of course, is a definite step towards the full
GW+DMFT scheme, but still one cannot say that there was the
summation of all "correlated" graphs in it which would require that
the W in GW part and $W_{imp}$ in the impurity (DMFT) part were the
same. Instead they were totally decoupled which makes it unclear of
what kind of diagrams from DMFT part were actually added to GW part.
Nevertheless, together with development of QMC solvers capable to
handle dynamical interactions,\cite{prb_72_035122,prl_97_076405} the
hope is growing that the full GW+DMFT scheme will eventually be
implemented.

There are some subtleties about the GW part as well. In its full sc
implementation the method is very time consuming which in part has
prevented its applications for the real materials. But recently a
very efficient implementation of it was
published\cite{prb_85_155129} where the most computationally
demanding parts (calculation of the polarizability P and the self
energy $\Sigma$) are performed in real space and Matsubara's time.
As a result, it became possible to successfully apply scGW to the
actinides Pu and Am and (earlier) to simple
sp-materials.\cite{prb_80_041103} However for the majority of
solids, scGW produces worse spectra than fast "one-shot" GW and this
is the other reason why scGW is not popular. The reason for the
failure of scGW with spectra may be traced as an extremely non
symmetrical "dressing" of the "initial" Green's function with self
energy insertions of GW-only form in the course of self-consistency
iterations and neglecting by vertex corrections.

The origin of the problem with the scGW method can also be
formulated in terms of the absence of the Z-factor
cancelation,\cite{prb_76_165106} which again happens because we
neglect by the vertex corrections. To resolve this problem Kotani
and Schilfgaarde\cite{prb_76_165106} devised a beautiful trick of
doing yet another approximation. They used quasiparticle (QP) part
of Green's function  $G_{QP}$ (instead of full Green's function $G$
calculated from Dyson's equation (DE)) to calculate $P$ and $\Sigma$
on every iteration till self-consistency. The trick is, that the
errors from the above two approximations (using $G_{QP}$ instead of
full $G$ and neglecting by vertex corrections) mostly cancel each
other out and as a result the QPscGW (self-consistent quasiparticle
GW) method usually gives much better spectra than full scGW. The
important fact is that QPscGW not just slightly improves the
one-shot GW description of sp-materials (which are good enough
already in the one-shot GW), but often gives reasonable results for
the materials with d- or f-electrons too,\cite{prl_96_226402} and
the method doesn't rely on a particular starting point. It is
totally self-consistent.

Quite naturally, the success of QPscGW with spectral properties (as
compared to the full scGW) has ignited the ideas to formulate
another approach - QPscGW+DMFT\cite{prl_109_237110,arx.1411.5180},
where one supposedly adds DMFT corrections to P and $\Sigma$ as in
GW+DMFT but uses QPscGW for the "big" iterations in the "weakly
correlated" part.

As it appears we have now two schemes proposed: GW+DMFT and
QPGW+DMFT. In this work I am doing an attempt to "estimate" what to
expect from the future implementation of the schemes. The analysis
strongly depends on the "separability" of weakly and strongly
correlated parts. I will assume here that they are perfectly
separable and the correlations in the GW part are really weak. In
this case, Z-factor in the GW part is close to 1, so that this part
is equally well described in both the GW and the QPGW
approximations. The difference correspondingly comes only from the
DMFT part. For this part I assume that we are able to solve it
exactly which means in particular that both sc conditions (for G and
W) are satisfied. For GW+DMFT, it means $G=G_{imp}$, $W=W_{imp}$,
and for QPGW+DMFT, it means $G_{QP}=G_{imp}$, $W_{QP}=W_{imp}$,
where following the arguments of work [\onlinecite{prb_76_165106}]
we have $G_{QP}\approx\frac{G}{Z}$. The exact solution of the
impurity problem also means that corresponding self energies can be
written in their exact diagrammatic forms: $\Sigma=-G\Gamma W$ and
$\Sigma_{QP}=-G_{QP}\Gamma W_{QP}$ with $\Gamma$ being the
three-point vertex function which is included exactly. But following
again the arguments in Ref.[\onlinecite{prb_76_165106}], it becomes
clear that in QPGW+DMFT case, i.e. when we include exact vertex and
continue to apply QP approximation for G, we will have a problem,
because the factor $1/Z$ appears from the vertex and it doesn't
cancel with the Z-factor in G as it happens in GW+DMFT. Basically it
means the violation of Ward Identity (WI) in the QPGW+DMFT scheme.
Thus, from this point of view, QPGW+DMFT is highly problematic and
its difficulties should grow up when the correlation strength grows
up, because the $1/Z$-factor increases.

The above simple argument against using the QPGW+DMFT scheme for
strongly correlated materials needs some numerical support and I
will provide it here using the exactly solvable two-site Hubbard
model. For this model I calculate exactly the three-point vertex
function and use it to calculate self-consistently the Green
functions G and $G_{QP}$ following two slightly different sc schemes
(the scheme on the left side is basically the Hedin's sc
equations\cite{pr_139_A796} but with the 3-point vertex
precalculated exactly)

\begin{align}\label{eq1}
\begin{array}{cc}
  P=G\Gamma_{Exact}G & P_{QP}=G_{QP}\Gamma_{Exact}G_{QP} \\
  W=U+UPW & W_{QP}=U+UP_{QP}W_{QP} \\
  \Sigma=-G\Gamma_{Exact}W & \Sigma_{QP}=-G_{QP}\Gamma_{Exact}W_{QP} \\
  G=G_{0}+G_{0}\Sigma G &
  G_{QP}=\frac{1}{Z}\frac{G_{0}}{1-G_{0}\Sigma_{QP}},
\end{array}
\end{align}
where $U$ is the bare interaction in the Hubbard model and $G_{0}$
if Green's function in Hartree approximation. In (\ref{eq1}) for QP
case I formally represented the quasiparticle approximation for G by
simply dividing the full Green function from Dyson's equation by the
Z-factor. This is for brevity. In fact I use the algorithm described
in Ref.[\onlinecite{prb_85_155129}] to construct $G_{QP}$.

It is obvious that at self-consistency the scheme on the left side
(I will call it $G\Gamma_{E}W$) is equivalent to the exact solution
of GW+DMFT equations whereas the right hand scheme (I will call it
$QPG\Gamma_{E}W$) is equivalent to the exact solution of the
QPGW+DMFT equations. It is also obvious that the $G\Gamma_{E}W$
scheme is exact by construction. I use it in this work to check the
numerical accuracy of 3-point vertex evaluation. The
$QPG\Gamma_{E}W$ scheme is approximate and below I will explore its
accuracy in different regimes of correlation strength for the 2-site
Hubbard model. Also I will directly relate the problems of the
$QPG\Gamma_{E}W$ scheme with the degree of the WI violation.

Another goal of the present work was to explore the possibility of
combining the GW and the QPGW methods with perturbative calculation
of the 3-point vertex function. To this end I will use again the
schemes similar to Eq.(\ref{eq1}) but with $\Gamma$ expanded to the
first order in W ($\Gamma_{1}$) instead of exact $\Gamma_{E}$. I
also will show how the two corresponding perturbation theory based
schemes ($G\Gamma_{1}W$ and $QPG\Gamma_{1}W$) behave in the
different regimes of parameters of the Hubbard model.

This paper begins with the formal presentation of the two-site
Hubbard model and the formulae used in the calculations (Section
\ref{HB2}). In Section~\ref{res} the results are presented and
discussed. Finally in Section~\ref{concl} the conclusions are given
and the future plans are outlined.

\section{Two-site Hubbard model}
\label{HB2}

The Hamiltonian of two-site Hubbard model as it is used in this work
is the following:

\begin{align}\label{eq2}
H=-t\sum_{i\neq
j,\sigma}c^{+}_{i\sigma}c_{j\sigma}+\frac{U}{2}\sum_{i}\sum_{\sigma\sigma'}c^{+}_{i\sigma}c^{+}_{i\sigma'}c_{i\sigma'}c_{i\sigma},
\end{align}
where $t$ and $U$ are the standard parameters of the Hubbard model,
$c$ and $c^{+}$ are the destruction and creation operators
correspondingly, indexes $i$ and $j$ belong to the sites (1 or 2),
and $\sigma,\sigma'$ are the spin indexes.

In this section, all equations which are used in the analysis of the
2-site Hubbard model are collected for references. First I provide
the energies and eigen vectors of the exact many-body states of the
model for different occupancies.  Then, the exact expressions for
Green function and density-density correlation function are given.
From them the exact self energy, polarizability, dielectric
function, and screened interaction can be calculated using standard
formulae. Next, two subsections provide the formulae for the exact
3-point vertex function and how it is used to evaluate the
corrections to the polarizability and to the self energy in sc
scheme (\ref{eq1}). Perturbation theory based equations for the
3-point vertex function are given next. Then I provide full and
reduced (long-wave and long-wave+static limits) expressions for the
Ward Identity which are used later in this paper. Finally the
formulae relevant to the evaluation of current three-point vertex
function are given. For brevity, the derivations of the formulae are
not provided, or only sketch of derivation is given. In this work,
the finite temperature framework is used so that in all equations
below the time-frequency arguments are Matsubara's time ($\tau$) and
Matsubara's frequencies ($\omega$ for fermion frequency and $\nu$
for boson frequency).

\subsection{Many-body states} \label{mbs}

In order to represent the many-body states of the model it is
convenient to introduce basis vectors $|abcd\rangle$ where all
entries are equal to 0 or 1 in accordance with the occupancies of
corresponding one-electron states. In this work first two
one-electron occupancies (a and b) correspond to spin-up and
spin-down one-electron states of the first site, and the third and
fourth (c and d) correspond to the second site. Many-body energies
and states below have two indexes: the upper one corresponds to the
full occupancy (0,1,2,3, or 4) of the system, and the lower one
distinguishes the states within the same occupancy.

For the full occupancy N equal zero, we have correspondingly:
\begin{align}\label{eq3}
E^{0}_{1}=0; \Psi^{0}_{1}=|0000\rangle,
\end{align}
for N=1
\begin{align}\label{eq3}
\begin{array}{c}
  E^{1}_{1}=-t; \Psi^{1}_{1}=\frac{1}{\sqrt{2}}(|1000\rangle+|0010\rangle) \\
  E^{1}_{2}=-t; \Psi^{1}_{2}=\frac{1}{\sqrt{2}}(|0100\rangle+|0001\rangle) \\
  E^{1}_{3}=t; \Psi^{1}_{3}=\frac{1}{\sqrt{2}}(|1000\rangle-|0010\rangle) \\
  E^{1}_{4}=t; \Psi^{1}_{4}=\frac{1}{\sqrt{2}}(|0100\rangle-|0001\rangle),
\end{array}
\end{align}
for N=2
\begin{align}\label{eq4}
\begin{array}{c}
  E^{2}_{1}=\frac{U-c}{2};\Psi^{2}_{1}=4t\frac{|1001\rangle-|0110\rangle}{\sqrt{a(c-U)}}+\frac{|1100\rangle+|0011\rangle}{a} \\
  E^{2}_{2}=0; \Psi^{2}_{2}=|1010\rangle \\
  E^{2}_{3}=0; \Psi^{2}_{3}=|0101\rangle \\
  E^{2}_{4}=0; \Psi^{2}_{4}=\frac{1}{\sqrt{2}}(|1001\rangle+|0110\rangle) \\
  E^{2}_{5}=U; \Psi^{2}_{5}=\frac{1}{\sqrt{2}}(|1100\rangle-|0011\rangle) \\
  E^{2}_{6}=\frac{U+c}{2};\Psi^{2}_{6}=4t\frac{|1001\rangle-|0110\rangle}{\sqrt{b(c+U)}}-\frac{|1100\rangle+|0011\rangle}{b},
\end{array}
\end{align}
with $a=\sqrt{2+\frac{32t^{2}}{(U-c)^{2}}}$,
$b=\sqrt{2+\frac{32t^{2}}{(U+c)^{2}}}$, and
$c=\sqrt{16t^{2}+U^{2}}$.

For N=3
\begin{align}\label{eq5}
\begin{array}{c}
  E^{3}_{1}=U-t; \Psi^{3}_{1}=\frac{1}{\sqrt{2}}(|1011\rangle-|1110\rangle) \\
  E^{3}_{2}=U-t; \Psi^{3}_{2}=\frac{1}{\sqrt{2}}(|0111\rangle-|1101\rangle) \\
  E^{3}_{3}=U+t; \Psi^{3}_{3}=\frac{1}{\sqrt{2}}(|1011\rangle+|1110\rangle) \\
  E^{3}_{4}=U+t; \Psi^{3}_{4}=\frac{1}{\sqrt{2}}(|0111\rangle+|1101\rangle),
\end{array}
\end{align}
and for N=4
\begin{align}\label{eq6}
E^{4}_{1}=2U; \Psi^{4}_{1}=|1111\rangle.
\end{align}

\subsection{The Partition function, Green's function and self energy} \label{GSig}

For convenience, first I introduce "shifted" many-body energies
$E'^{N}_{n}=E^{N}_{n}-\mu N$, with $\mu$ being the chemical
potential. Then I renormalize them, defying the minimal $E'_{min}$
among them and subtracting it $E''^{N}_{n}=E'^{N}_{n}-E'_{min}$.
This also factorizes the partition function:
\begin{align}\label{eq7}
Z(\mu)&=\sum_{nN}e^{-\beta E'^{N}_{n}}=e^{-\beta
E'_{min}}\sum_{nN}e^{-\beta E''^{N}_{n}}\nonumber\\&=e^{-\beta
E'_{min}}Z'(\mu),
\end{align}
where $\beta$ is the inverse temperature.

It is clear that now in every Gibbs average one can use $E'';Z'$
instead of $E';Z$ which is numerically more stable (big numbers have
been subtracted).

For exact Green's function, one obtains through the standard
spectral decomposition:

\begin{align}\label{eq8}
G^{\sigma}_{ij}(\omega)&=\frac{1}{Z'}\sum_{N}\sum_{n\in N}\sum_{m\in
N+1}\frac{e^{-\beta E''^{N+1}_{m}}+e^{-\beta
E''^{N}_{n}}}{i\omega-E''_{mN+1}+E''_{nN}}\nonumber\\&\times \langle
\Psi^{N}_{n}|c_{i\sigma}|\Psi^{N+1}_{m}\rangle \langle
\Psi^{N+1}_{m} |c^{+}_{j\sigma}|\Psi^{N}_{n}\rangle.
\end{align}

The exact self energy is obtained by inversion of the Dyson's
equation:
\begin{align}\label{eq9}
\Sigma^{\sigma}_{ij}(\omega)=G^{^{-1}\sigma}_{0,ij}(\omega)-G^{^{-1}\sigma}_{ij}(\omega),
\end{align}
where Green's function in Hartree approximation
$G^{^{-1}\sigma}_{0,ij}(\omega)=(i\omega+\mu-U\rho_{i})\delta_{ij}+t(1-\delta_{ij})$
is used ($\rho_{i}$ is the occupancy ("density") of the site $i$).

\subsection{Response function, polarizability, dielectric function, and W} \label{Chi_W}

The exact two-point density-density correlation function is also
obtained through the spectral decomposition

\begin{align}\label{eq10}
\chi^{dd}_{ij}(\nu)&=\frac{1}{Z'}\sum_{N}\sum_{n\in N}\sum_{m\in
N}\frac{e^{-\beta E''^{N}_{m}}-e^{-\beta
E''^{N}_{n}}}{i\nu+E''_{nN}-E''_{mN}}\nonumber\\&\times \langle
\Psi^{N}_{n} |\hat{\rho}_{i}|\Psi^{N}_{m}\rangle \langle
\Psi^{N}_{m} |\hat{\rho}_{j}|\Psi^{N}_{n}\rangle,
\end{align}
with density operator
$\hat{\rho}_{i}=\sum_{\sigma}c^{+}_{i\sigma}c_{i\sigma}$. It is
convenient to define also the density-density response function

\begin{align}\label{eq11}
R^{dd}_{ij}(\nu)=\beta
\delta_{\nu0}\rho_{i}\rho_{j}-\chi^{dd}_{ij}(\nu).
\end{align}
After that one can find the density-density dielectric function
\begin{align}\label{eq12}
\epsilon^{^{-1}dd}_{ij}(\nu)=\delta_{ij}+UR^{dd}_{ij}(\nu),
\end{align}
the density-density polarizability
\begin{align}\label{eq13}
P^{dd}_{ij}(\nu)=\sum_{k}R^{dd}_{ik}(\nu)\epsilon^{dd}_{kj}(\nu),
\end{align}
and the screened interaction
\begin{align}\label{eq14}
W_{ij}(\nu)=U\delta_{ij}+U^{2}R^{dd}_{ij}(\nu).
\end{align}

The response function, the dielectric function, the polarizability,
and the screened interaction calculated using the formulae
(\ref{eq11})-(\ref{eq14}) from the exact correlation function are by
construction exact and can be compared with the corresponding
quantities obtained using the PT.

\subsection{Exact 3-point vertex function in density channel} \label{Gamma_exact}

To find the exact 3-point density vertex function I first calculate
the following three-point correlation function:
$\chi^{\sigma,d}_{ijk}(\tau;\tau')=\langle
c_{i\sigma}(\tau)c^{+}_{j\sigma}(\tau')\hat{\rho}_{k}(0)\rangle$
with $\tau,\tau'$ being Matsubara's times and 'd' meaning 'density'.
In the site-frequency domain its spectral decomposition reads as the
following:

\begin{widetext}
\begin{align}\label{eq15}
\chi^{\sigma,d}_{ijk}&(\omega;\nu)=\frac{1}{Z'}\sum_{N}\Big\{\nonumber\\&-\sum_{p\in
N}\sum_{n\in N+1}\frac{\langle \Psi^{N+1}_{n}
|c^{+}_{j\sigma}|\Psi^{N}_{p}\rangle}{i(\omega-\nu)+E''^{N}_{p}-E''^{N+1}_{n}}\sum_{m\in
N+1}\langle \Psi^{N}_{p} |c_{i\sigma}|\Psi^{N+1}_{m}\rangle\langle
\Psi^{N+1}_{m} |\hat{\rho}_{k}|\Psi^{N+1}_{n}\rangle\frac{e^{-\beta
E''^{N+1}_{m}}-e^{-\beta
E''^{N+1}_{n}}}{i\nu+E''^{N+1}_{n}-E''^{N+1}_{m}}\nonumber\\&+\sum_{p\in
N+1}\sum_{m\in N}\frac{\langle \Psi^{N+1}_{p}
|c^{+}_{j\sigma}|\Psi^{N}_{m}\rangle}{-i(\omega-\nu)+E''^{N+1}_{p}-E''^{N}_{m}}\sum_{n\in
N}\langle \Psi^{N}_{n} |c_{i\sigma}|\Psi^{N+1}_{p}\rangle\langle
\Psi^{N}_{m} |\hat{\rho}_{k}|\Psi^{N}_{n}\rangle\frac{e^{-\beta
E''^{N}_{m}}-e^{-\beta
E''^{N}_{n}}}{i\nu+E''^{N}_{n}-E''^{N}_{m}}\nonumber\\&+\sum_{p\in
N}\sum_{n\in N+1}\frac{\langle \Psi^{N+1}_{n}
|c^{+}_{j\sigma}|\Psi^{N}_{p}\rangle}{i(\omega-\nu)+E''^{N}_{p}-E''^{N+1}_{n}}\sum_{m\in
N+1}\langle \Psi^{N}_{p} |c_{i\sigma}|\Psi^{N+1}_{m}\rangle\langle
\Psi^{N+1}_{m} |\hat{\rho}_{k}|\Psi^{N+1}_{n}\rangle\frac{e^{-\beta
E''^{N+1}_{m}}+e^{-\beta
E''^{N}_{p}}}{i\omega+E''^{N}_{p}-E''^{N+1}_{m}}\nonumber\\&+\sum_{p\in
N+1}\sum_{m\in N}\frac{\langle \Psi^{N+1}_{p}
|c^{+}_{j\sigma}|\Psi^{N}_{m}\rangle}{-i(\omega-\nu)+E''^{N+1}_{p}-E''^{N}_{m}}\sum_{n\in
N}\langle \Psi^{N}_{n} |c_{i\sigma}|\Psi^{N+1}_{p}\rangle\langle
\Psi^{N}_{m} |\hat{\rho}_{k}|\Psi^{N}_{n}\rangle\frac{e^{-\beta
E''^{N+1}_{p}}+e^{-\beta
E''^{N}_{n}}}{i\omega+E''^{N}_{n}-E''^{N+1}_{p}}\Big\}.
\end{align}
\end{widetext}

After that the three-point density response function is calculated

\begin{align}\label{eq16}
R^{\sigma,d}_{ijk}(\omega;\nu)=\frac{\delta
G^{\sigma}_{ij}(\omega)}{\delta \phi_{k}(\nu)}=\beta
\delta_{\nu0}G^{\sigma}_{ij}(\omega)\rho_{k}+\chi^{\sigma,d}_{ijk}(\omega;\nu),
\end{align}
where I have indicated that the three-point response function is
defined as the functional derivative of Green's function with
respect to the external perturbing field $\phi_{k}(\nu)$. The
"screened" 3-point density vertex function
$\gamma^{\sigma,d}_{ijk}(\omega;\nu)$ is defined as the functional
derivative of the inverse Green's function with respect to the
external perturbing field $\phi_{k}(\nu)$ and correspondingly can be
related to the above defined 3-point response function

\begin{align}\label{eq17}
\gamma^{\sigma,d}_{ijk}(\omega;\nu)&=-\frac{\delta
G^{^{-1}\sigma}_{ij}(\omega)}{\delta \phi_{k}(\nu)}\nonumber\\&
=\sum_{lt}G^{^{-1}\sigma}_{il}(\omega)R^{\sigma,d}_{ltk}(\omega;\nu)G^{^{-1}\sigma}_{tj}(\omega-\nu).
\end{align}

Finally the three-point vertex function entering the Eq.(\ref{eq1})
("bare" three-point vertex) is defined as the functional derivative
of the inverse Green's function with respect to the total field
$\Phi_{k}(\nu)$ (perturbing external plus induced internal) and is
related to the "screened" vertex through the density-density
dielectric matrix (this equation is the density-density part of the
more general equation (\ref{eq31i}))

\begin{align}\label{eq18}
\Gamma^{\sigma,d}_{ijk}(\omega;\nu)&=-\frac{\delta
G^{^{-1}\sigma}_{ij}(\omega)}{\delta
\Phi_{k}(\nu)}\nonumber\\&=\sum_{l}\gamma^{\sigma,d}_{ijl}(\omega;\nu)\epsilon^{dd}_{lk}(\nu).
\end{align}

\subsection{Vertex corrected polarizability and self energy} \label{dP_dSig}

The vertex-corrected density-density polarizability and the self
energy entering the equations (\ref{eq1}) are calculated as the
following

\begin{align}\label{eq19}
P^{dd}_{ij}(\nu)=\frac{1}{\beta}\sum_{\omega}\sum_{\sigma}\sum_{kl}
G^{\sigma}_{ik}(\omega)\Gamma^{\sigma,d}_{klj}(\omega;\nu)G^{\sigma}_{li}(\omega-\nu),
\end{align}
and
\begin{align}\label{eq20}
\Sigma^{\sigma}_{ij}(\omega)=-\frac{1}{\beta}\sum_{\nu}\sum_{kl}
G^{\sigma}_{ik}(\omega+\nu)\Gamma^{\sigma,d}_{kjl}(\omega+\nu;\nu)W_{li}(\nu).
\end{align}

The formulae (\ref{eq19}) and (\ref{eq20}) are used also when the
three-point vertex is obtained within the perturbation theory.

\subsection{3-point vertex function from Perturbation Theory in density channel} \label{Gamma_pert}

The first order (in W) term of the perturbation theory for the
three-point density vertex function is

\begin{align}\label{eq21}
\Gamma^{\sigma,d}_{ijk}(\tau;\tau')=-W_{ji}(\tau'-\tau)G^{\sigma}_{ik}(\tau)G^{\sigma}_{kj}(-\tau').
\end{align}

In the frequency representation it can be conveniently evaluated as
the following
\begin{align}\label{eq22}
\Gamma^{\sigma,d}_{ijk}(\omega;\nu)=-\int d\tau
W_{ji}(\tau)\frac{1}{\beta}\sum_{\omega}
e^{-i\omega\tau}G^{\sigma}_{ik}(\omega)G^{\sigma}_{kj}(\omega-\nu).
\end{align}

\subsection{Ward Identities} \label{WIhm}

In order to write down the Ward Identities one needs to specify the
current operator. It is convenient to introduce it through the
substitution for the kinetic part of the Hamiltonian

\begin{align}\label{eq22a}
H'=-t\sum_{i\sigma}c^{+}_{i\sigma}e^{i(A_{\tilde{i}}-A_{i})}c_{\tilde{i}\sigma}
+\frac{U}{2}\sum_{i}\sum_{\sigma\sigma'}c^{+}_{i\sigma}c^{+}_{i\sigma'}c_{i\sigma'}c_{i\sigma},
\end{align}
where $A_{i}$ is the vector potential and the following convention
for the sites was adopted: $\tilde{1}=2;\tilde{2}=1$. With the above
definition the current operator for the model reads as the following

\begin{align}\label{eq23}
\hat{J}_{i}=-\frac{\delta H'}{\delta
A_{i}}=it\sum_{\sigma}\big\{c^{+}_{i\sigma}c_{\tilde{i}\sigma}-c^{+}_{\tilde{i}\sigma}c_{i\sigma}\big\},
\end{align}
with $\hat{J}_{\tilde{i}}=-\hat{J}_{i}$. From the equation of motion
for the density operator one calculates

\begin{align}\label{eq24}
\frac{\partial \hat{\rho}_{i}}{\partial\tau}=-i\hat{J}_{i},
\end{align}
which is the continuity equation for the model. Relating the
"screened" 3-point current vertex function with the corresponding
3-point response function (similarly to the density case and in the
space-time coordinates for brevity; 'c' goes for 'current'))

\begin{align}\label{eq25}
G^{\sigma}(14)\gamma^{\sigma,c}(453)G^{\sigma}(52)&=G^{\sigma}(12)J(3)\nonumber\\&+\langle
c_{\sigma}(1)c^{+}_{\sigma}(2)\hat{J}(3)\rangle,
\end{align}
one explicitly calculates the time derivatives and using also the
continuity equation (\ref{eq24}) one obtains
\begin{align}\label{eq26}
G^{\sigma}&(14)\Big\{\frac{\partial}{\partial\tau_{3}}\gamma^{\sigma,d}(453)+i\gamma^{\sigma,c}(453)\Big\}G^{\sigma}(52)\nonumber\\&
=\frac{\partial}{\partial\tau_{3}}\langle
c_{\sigma}(1)c^{+}_{\sigma}(2)\hat{\rho}(3)\rangle+i\langle
c_{\sigma}(1)c^{+}_{\sigma}(2)\hat{J}(3)\rangle\nonumber\\&
=G^{\sigma}(12)\Big\{\delta(13)-\delta(23)\Big\},
\end{align}
which is equivalent to

\begin{align}\label{eq27}
\frac{\partial}{\partial\tau_{3}}\gamma^{\sigma,d}(123)+i\gamma^{\sigma,c}(123)
=G^{^{-1}\sigma}(12)\Big\{\delta(23)-\delta(13)\Big\}.
\end{align}

In the site-frequency representation (\ref{eq27}) reads as the
following

\begin{align}\label{eq28}
i\Big\{\nu\gamma^{\sigma,d}_{ijk}(\omega;\nu)&+\gamma^{\sigma,c}_{ijk}(\omega;\nu)\Big\}
\nonumber\\&=G^{^{-1}\sigma}_{ij}(\omega)\delta_{jk}-G^{^{-1}\sigma}_{ij}(\omega-\nu)\delta_{ik}.
\end{align}

The equation (\ref{eq28}) can be simplified by removing the
Hartree-Fock contribution on the both sides of it. For the vertices,
one obtains in the Hartree-Fock approximation

\begin{align}\label{eq29}
\gamma^{\sigma,d}_{ijk}(\omega;\nu)=\delta_{ik}\delta_{jk},
\end{align}
and
\begin{align}\label{eq30}
\gamma^{\sigma,c}_{ijk}(\omega;\nu)=it\Big\{\delta_{ik}-\delta_{jk}\Big\}.
\end{align}

For the schemes with full sc (without the QP approximation) the
removing of the Hartree-Fock contribution also on the right side of
(\ref{eq28}) through Dyson's equation gives

\begin{align}\label{eq31}
i\Big\{\nu\triangle\gamma^{\sigma,d}_{ijk}(\omega;\nu)&+\triangle\gamma^{\sigma,c}_{ijk}(\omega;\nu)\Big\}
\nonumber\\&=\Sigma^{c,\sigma}_{ij}(\omega-\nu)\delta_{ik}-\Sigma^{c,\sigma}_{ij}(\omega)\delta_{jk},
\end{align}
where $\triangle\gamma$ means the "screened" vertex part beyond the
Hartree-Fock approximation, and the $\Sigma^{c}$ is the correlation
(frequency dependent) part of the self energy. For the QP-based
schemes the Dyson equation is not satisfied and instead of
(\ref{eq31}) one has
\begin{align}\label{eq31a}
i\Big\{\nu\triangle\gamma^{\sigma,d}_{ijk}(\omega;\nu)&+\triangle\gamma^{\sigma,c}_{ijk}(\omega;\nu)\Big\}
\nonumber\\&=\Big\{H^{HF}_{ij}-H^{QP}_{ij}\Big\}[\delta_{ik}-\delta_{jk}],
\end{align}
where the static effective Hamiltonians $H^{HF}_{ij}$ and
$H^{QP}_{ij}$ (correspondingly in the Hartree-Fock and in the QP
approximation) were introduced:

\begin{align}\label{eq31b}
H^{HF}_{ij}=-t(1-\delta_{ij})+\delta_{ij}(V^{H}_{i}+\Sigma^{x}_{i}),
\end{align}
with $V^{H}_{i}$ and $\Sigma^{x}_{i}$ being the Hartree potential
and the exchange part of the self energy correspondingly, and
\begin{align}\label{eq31c}
H^{QP}_{ij}=\mu[\delta_{ij}-Z_{ij}]+\sum_{kl}Z^{1/2}_{ik}[H^{HF}_{kl}+\Sigma^{c}_{kl}(0)]Z^{1/2}_{lj},
\end{align}
where $Z_{ij}$ is the renormalization factor, and the
$\Sigma^{c}_{kl}(0)$ stands for the correlation part of the self
energy at zero frequency. When the effect of correlations is
neglected the renormalization factor becomes equal to 1, and the
correlation part of the self energy becomes zero, which means that
in this case $H^{QP}=H^{HF}$.

The equations (\ref{eq31}) and (\ref{eq31a}) are used in this work
to evaluate the deviations from the full Ward Identity for different
approximate methods. I also use two reduced forms of the WI in the
present study: the long-wave limit and the long-wave+static limit of
the WI. The long-wave limit of the WI for the two-site Hubbard model
consists in the summation over the index $k$ in the equations
(\ref{eq31}) and (\ref{eq31a}) which correspondingly become (the
current vertex disappears after the summation)

\begin{align}\label{eq31d}
i\nu\sum_{k}\triangle\gamma^{\sigma,d}_{ijk}(\omega;\nu)=\Sigma^{c,\sigma}_{ij}(\omega-\nu)-\Sigma^{c,\sigma}_{ij}(\omega),
\end{align}
and
\begin{align}\label{eq31e}
i\nu\sum_{k}\triangle\gamma^{\sigma,d}_{ijk}(\omega;\nu)=0.
\end{align}

From (\ref{eq31e}) one can see that in order to satisfy the
long-wave limit of the WI in the QP-based approximations one has to
neglect by vertex corrections altogether.

The long-wave+static limit of the WI consists in taking the limit
($\nu\rightarrow0$) in the equations (\ref{eq31d}) and
(\ref{eq31e}):

\begin{align}\label{eq31f}
\sum_{k}\triangle\gamma^{\sigma,d}_{ijk}(\omega;\nu=0)=\lim_{\nu\rightarrow0}
\frac{\Sigma^{c,\sigma}_{ij}(\omega-\nu)-\Sigma^{c,\sigma}_{ij}(\omega)}{i\nu},
\end{align}
and
\begin{align}\label{eq31g}
\sum_{k}\triangle\gamma^{\sigma,d}_{ijk}(\omega;\nu=0)=0.
\end{align}

The arguments supporting the quasiparticle approximation in
Ref.[\onlinecite{prb_76_165106}] are based on the long wave limit of
the Ward Identity. It is clear from the above consideration that in
the QPGW approximation (without the vertex corrections) the
corresponding limit of the WI is satisfied exactly.

\subsection{3-point current vertex function} \label{cur_vrt}

In order to apply the full WI one needs the "screened" current
vertex function $\gamma^{\sigma,c}_{ijk}(\omega;\nu)$. In this work
both the exact and the PT-based vertex functions are used. The exact
one is evaluated following the formulae (\ref{eq15})-(\ref{eq17})
with the replacement ($\hat{\rho}\rightarrow\hat{J}$) in the
Eq.(\ref{eq15}). The PT-based vertices are calculated following the
scheme outlined below.

The "bare" vertex functions are related to the "screened" ones
through the full (density-current) dielectric function via the
following equation which is the generalization of Eq.(\ref{eq18})

\begin{align}\label{eq31i}
\gamma^{\sigma,I}_{ijk}(\omega;\nu)=\sum_{J}\sum_{l}\Gamma^{\sigma,J}_{ijl}(\omega;\nu)\epsilon^{^{-1}JI}_{lk}(\nu),
\end{align}
where both $I$ and $J$ now run over the indices $d$ (density) and
$c$ (current). The non-perturbed Hamiltonian has no vector
potentials so that the full dielectric matrix has the form
(considering the first index as the density, and the second as the
current)
\begin{align}\label{eq31j}
\epsilon=\left(
           \begin{array}{cc}
             \epsilon^{dd} & \epsilon^{dc} \\
             0 & 1 \\
           \end{array}
         \right),
\end{align}
and correspondingly its inverse
\begin{align}\label{eq31k}
\epsilon^{-1}=\left(
           \begin{array}{cc}
             \epsilon^{^{-1}dd} & -\epsilon^{^{-1}dd}\epsilon^{dc} \\
             0 & 1 \\
           \end{array}
         \right).
\end{align}

Thus, the bare current vertex can be evaluated as the following

\begin{align}\label{eq31l}
\gamma^{\sigma,c}_{ijk}(\omega;\nu)&=\Gamma^{\sigma,c}_{ijl}(\omega;\nu)\nonumber\\&-\sum_{lm}\Gamma^{\sigma,d}_{ijl}(\omega;\nu)
\epsilon^{^{-1}dd}_{lm}(\nu)\epsilon^{dc}_{mk}(\nu).
\end{align}

The subtraction of the Hartree-Fock contribution (\ref{eq30}) leaves
us with the expression

\begin{align}\label{eq31m}
\triangle\gamma^{\sigma,c}_{ijk}(\omega;\nu)&=\triangle\Gamma^{\sigma,c}_{ijl}(\omega;\nu)\nonumber\\&
-\sum_{lm}[\delta_{il}\delta_{jl}+\triangle\Gamma^{\sigma,d}_{ijl}(\omega;\nu)]
\epsilon^{^{-1}dd}_{lm}(\nu)\epsilon^{dc}_{mk}(\nu).
\end{align}

In the above expression the missing components are the non-trivial
part of the current vertex
$\triangle\Gamma^{\sigma,c}_{ijl}(\omega;\nu)$ and the
density-current dielectric matrix $\epsilon^{dc}_{mk}(\nu)$. The
first one is evaluated similar to the equation (\ref{eq22}) for the
density vertex:

\begin{align}\label{eq31n}
\triangle&\Gamma^{\sigma,c}_{ijk}(\omega;\nu)=-it\int d\tau
W_{ji}(\tau)\frac{1}{\beta}\sum_{\omega}
e^{-i\omega\tau}\nonumber\\&\times\Big\{G^{\sigma}_{ik}(\omega)G^{\sigma}_{\tilde{k}j}(\omega-\nu)
-G^{\sigma}_{i\tilde{k}}(\omega)G^{\sigma}_{kj}(\omega-\nu)\Big\}.
\end{align}

The second one is defined by the density-current polarizability
$P^{dc}$

\begin{align}\label{eq31o}
\epsilon^{dc}_{ij}(\nu)=-UP^{dc}_{ij}(\nu),
\end{align}
which in its turn is evaluated using the current vertex
(\ref{eq31n}):

\begin{align}\label{eq31p}
&P^{dc}_{ij}(\nu)\nonumber\\&=it\int d\tau
e^{i\nu\tau}\sum_{\sigma}\Big\{G^{\sigma}_{ij}(\tau)G^{\sigma}_{\tilde{j}i}(-\tau)
-G^{\sigma}_{i\tilde{j}}(\tau)G^{\sigma}_{ji}(-\tau)\Big\}\nonumber\\&+\frac{1}{\beta}\sum_{\omega}\sum_{\sigma}\sum_{kl}
G^{\sigma}_{ik}(\omega)\triangle\Gamma^{\sigma,c}_{klj}(\omega;\nu)G^{\sigma}_{li}(\omega-\nu).
\end{align}

\subsection{Internal energy} \label{Eint}

The exact internal energy is evaluated directly as the average value
of the Hamiltonian. In the spectral representation it reads

\begin{align}\label{eq32}
E=\frac{1}{Z'}\sum_{nN}E''^{N}_{n}e^{-\beta E''^{N}_{n}}.
\end{align}

To evaluate the internal energy in the perturbation theory based
methods the following formula is used

\begin{align}\label{eq33}
E=&t\sum_{i\neq
j,\sigma}G^{\sigma}_{ji}(\tau=\beta)+\frac{U}{2}\sum_{i}\sum_{\sigma\sigma'}\rho_{i\sigma}\rho_{i\sigma'}
\nonumber\\&-\frac{1}{2}\sum_{i}\sum_{\sigma}\Sigma^{x,\sigma}_{ii}G^{\sigma}_{ii}(\tau=\beta)\nonumber\\&
+\frac{1}{2\beta}\sum_{i\sigma}\sum_{\omega}\Sigma^{c,\sigma}_{ij}(\omega)G^{\sigma}_{ji}(\omega),
\end{align}

which is based on the Galitskii-Migdal expression for the
exchange-correlation energy.

\section{Results} \label{res}

\begin{figure}[t]
\centering
\includegraphics[width=9.0 cm]{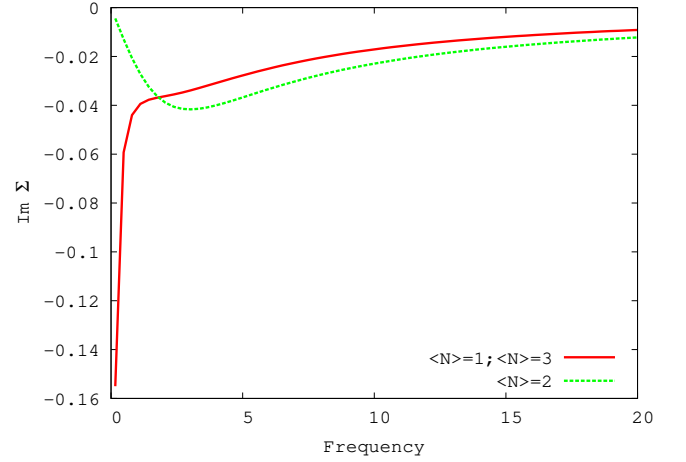}
\caption{(Color online) Imaginary part of the on-site self energy as
a function of frequency for different average occupancies for U=1.}
\label{sig_1_123}
\end{figure}

The two-site Hubbard model is studied here with the value of
parameter $t$ fixed and equal to 1. The temperature was also fixed
at T=0.05$t$. Thus, only the parameter U was changing. The
half-filling ($\langle N\rangle=2$) case has been considered. Such a
choice for the occupancy has been guided mostly by the fact that
when one steps aside from the half-filling the correlations in the
model quickly become unmanageable for the QP-based approaches. It is
seen in the Fig.\ref{sig_1_123} where the imaginary part of the
on-site self energy is plotted as a function of Matsubara's
frequency for the average occupancies 1, 2, and 3.

\clearpage

\begin{figure}[t]
\centering
\includegraphics[width=9.0 cm]{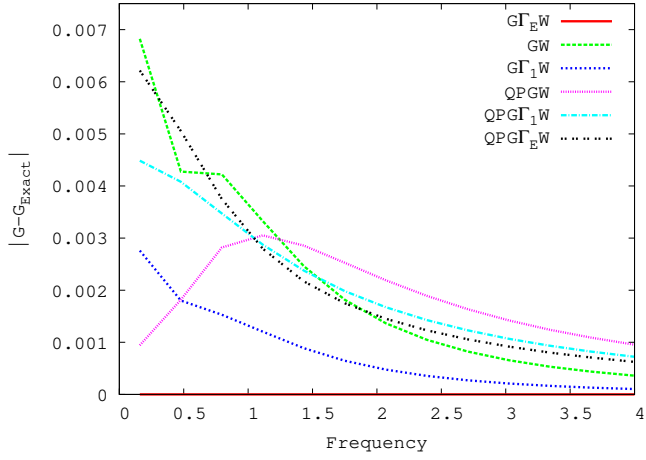}
\caption{(Color online) Errors in Green's function for U=0.5.}
\label{g_05_2}
\end{figure}

\begin{figure}[t]
\centering
\includegraphics[width=9.0 cm]{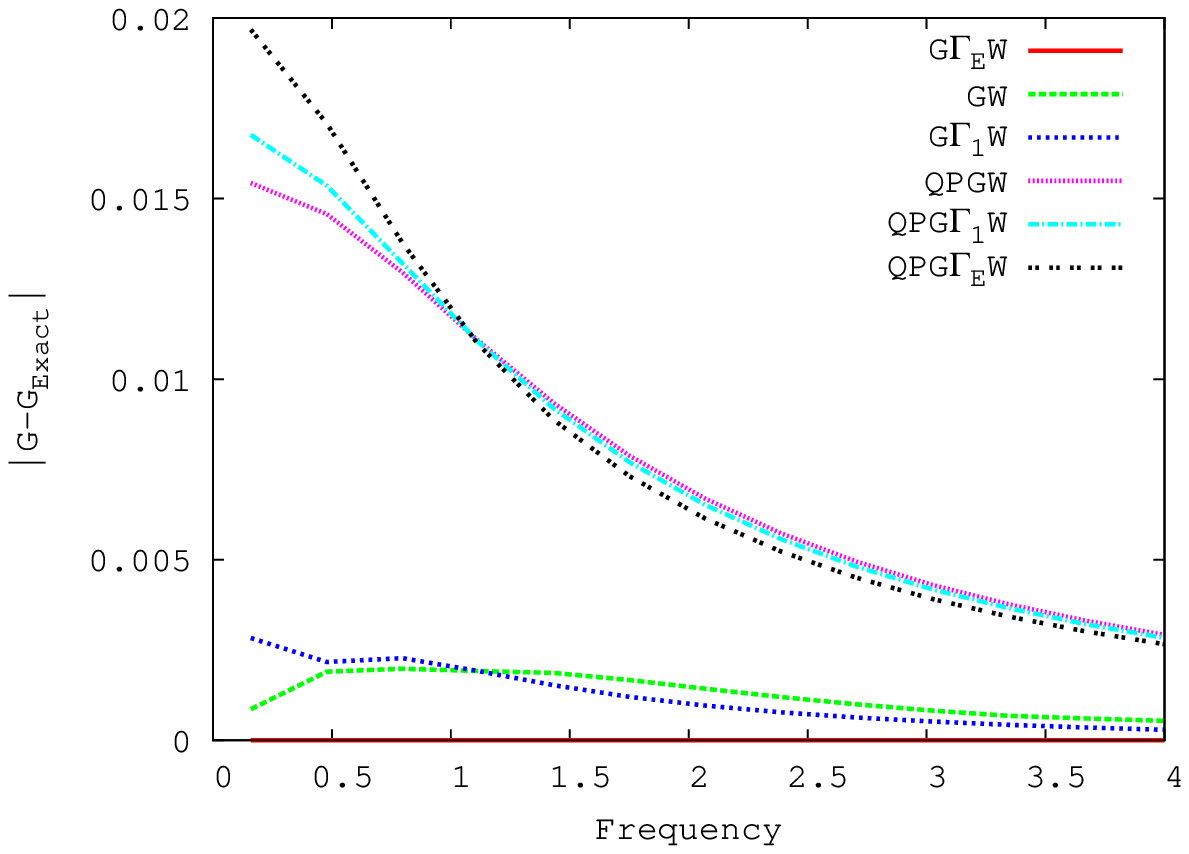}
\caption{(Color online) Errors in Green's function for U=1.}
\label{g_1_2}
\end{figure}

\begin{figure}[t]
\centering
\includegraphics[width=9.0 cm]{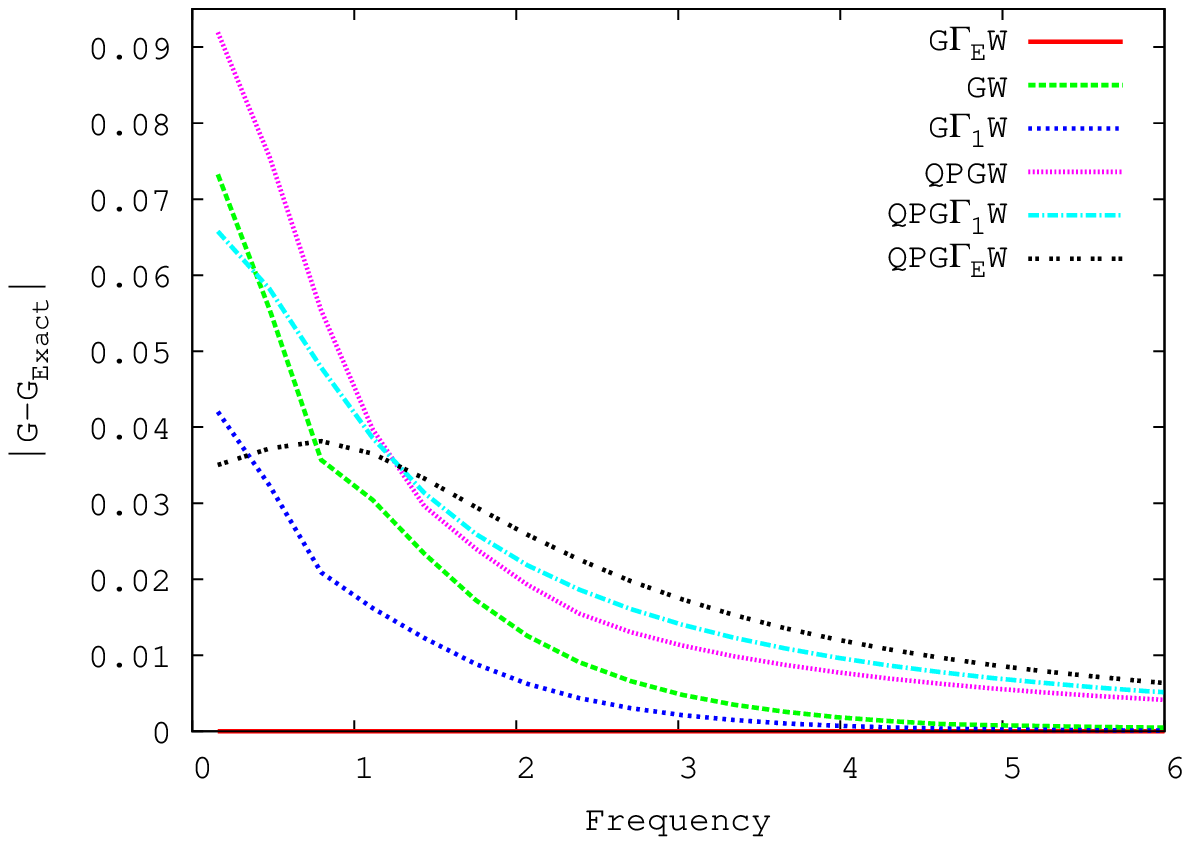}
\caption{(Color online) Errors in Green's function for U=2.}
\label{g_2_2}
\end{figure}

\begin{figure}[t]
\centering
\includegraphics[width=9.0 cm]{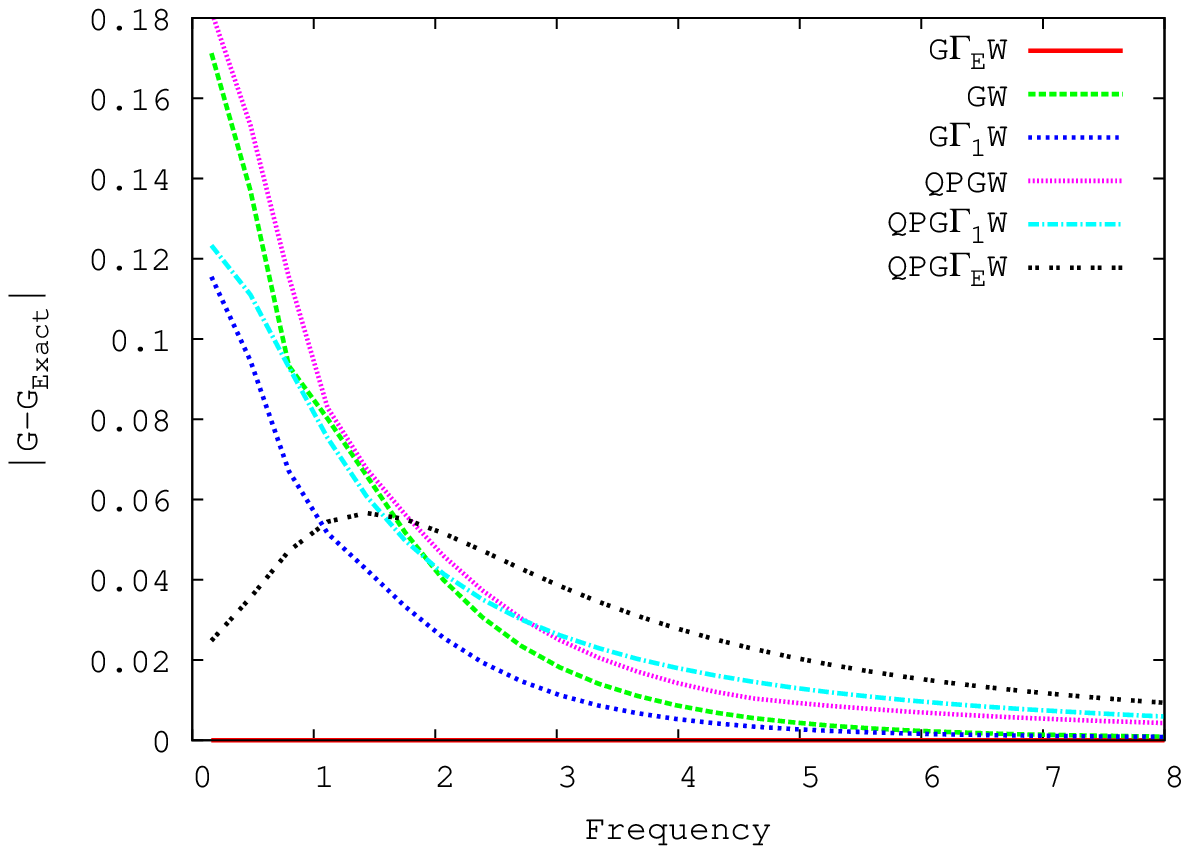}
\caption{(Color online) Errors in Green's function for U=3.}
\label{g_3_2}
\end{figure}

\begin{figure}[t]
\centering
\includegraphics[width=9.0 cm]{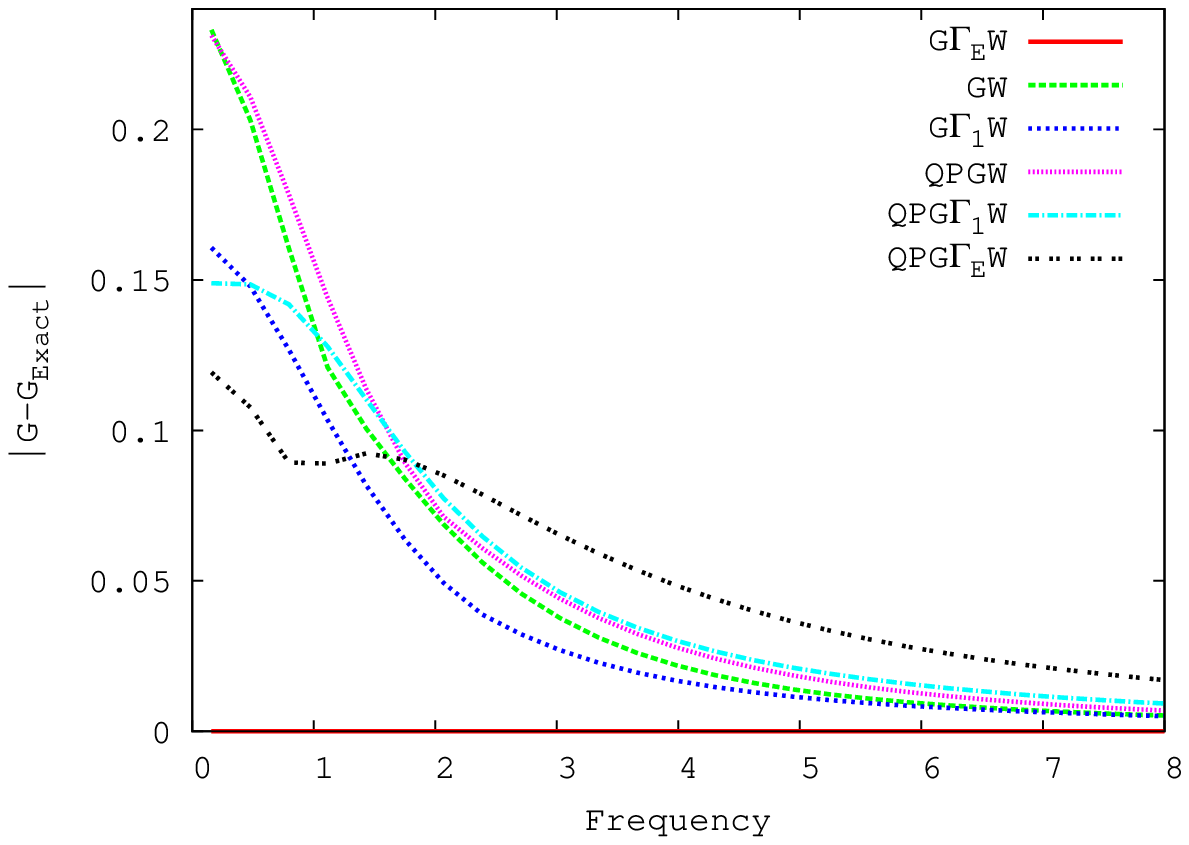}
\caption{(Color online) Errors in Green's function for U=4.}
\label{g_4_2}
\end{figure}

\begin{figure}[t]
\centering
\includegraphics[width=9.0 cm]{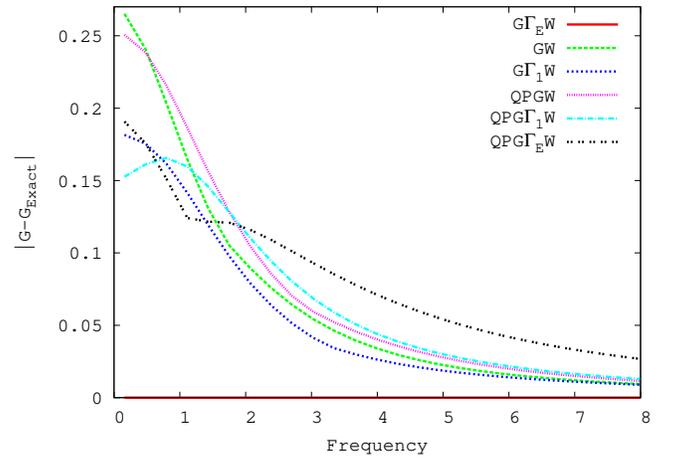}
\caption{(Color online) Errors in Green's function for U=5.}
\label{g_5_2}
\end{figure}

\clearpage

The strong downturn in the function for the occupancies 1 and 3 at
the small frequencies makes the linearization in the self energy
pertinent to the QP approximations highly inaccurate. At the half
filling however, the model shows slow increasing in the correlation
strength with the increasing of U and, correspondingly, is
convenient for the comparative studies. The calculations have been
performed for different values of U which were increased till the
methods based on the QP approximation began to fail seriously.

I compare the exact results with the results obtained with the GW,
the G$\Gamma_{1}$W, the G$\Gamma_{E}$W, the QPGW, QPG$\Gamma_{1}$W,
and the QPG$\Gamma_{E}$W methods, where $\Gamma_{1}$ and
$\Gamma_{E}$ stand for the first order (in W) 3-point vertex and the
exact 3-point vertex correspondingly. At the full self-consistency
G$\Gamma_{E}$W reproduces the exact results, so this approach was
used basically only as a mean to adjust (by comparing with the exact
result) the calculational parameters (such as the number of
Matsubara's frequencies included in the internal summations and the
density of mesh on the interval $[0:\beta]$ for the
$\tau$-integrations. In addition to the Green function which serves
in this work as basic representation quantity for the comparisons,
the analysis of the Ward Identities fulfillment has been performed
and was used as an indicator of the accuracy of the approximations.
Besides, the internal energy has been evaluated and its accuracy was
related to the errors in the calculated G and to the deviations from
the Ward Identities.

\begin{figure}[t]
\centering
\includegraphics[width=9.0 cm]{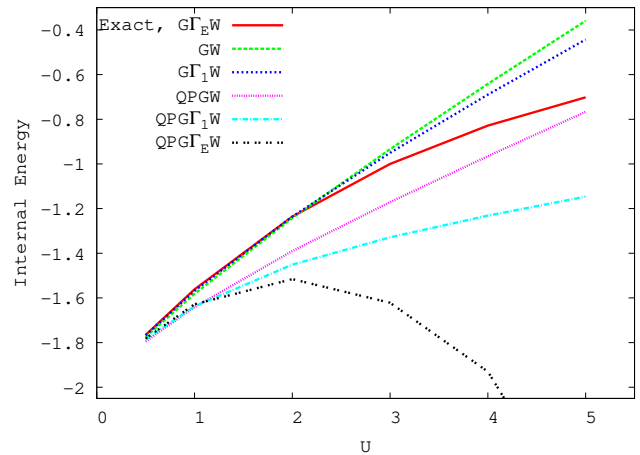}
\caption{(Color online) Internal energy as a function of U
parameter.} \label{E_2}
\end{figure}

\begin{figure}[b]
\centering
\includegraphics[width=9.0 cm]{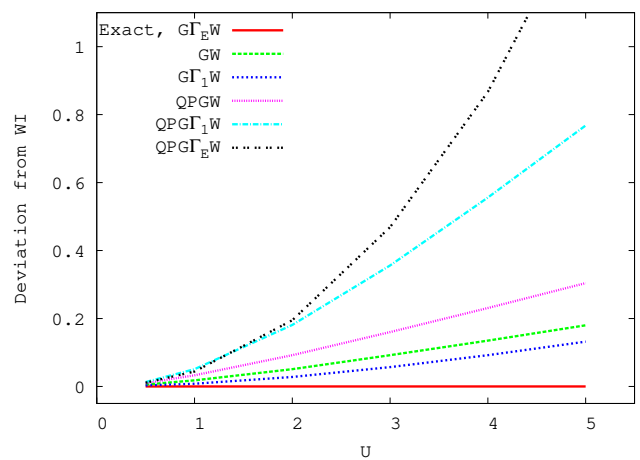}
\caption{(Color online) Full Ward Identity average violation as a
function of U parameter.} \label{F_WI_2}
\end{figure}

In the figures \ref{g_05_2}-\ref{g_5_2} the absolute error in the
calculated Green's function is shown as a function of the
Matsubara's frequency for the different values of U. The comparison
is being made between the exact result and the results obtained with
the approximate methods. The internal energy as a function of U is
presented in the Fig.\ref{E_2}. The figure \ref{F_WI_2} shows the
relative violation of the full WI (Eqs.\ref{eq31} and \ref{eq31a})
in different approximate methods as a function of U. The following
form of the average deviation from the identity has been used

\begin{align}\label{eq40}
\frac{\sum_{\omega,\nu}\sum_{ijk}\Big\|(LHS)_{ijk}(\omega;\nu)-(RHS)_{ijk}(\omega;\nu)\Big\|}{\sum_{\omega,\nu}\sum_{ijk}1},
\end{align}
where $(LHS)$ and $(RHS)$ are the left- and the right-hand sides in
(\ref{eq31},\ref{eq31a}) correspondingly. The 3-point vertex
pertinent to the specific approximation (i.e. for instance the
vertex $\triangle\gamma$ is zero in the GW and the QPGW) was used to
evaluate the deviation from the WI in all approximate schemes, and
the exact vertex was used in G$\Gamma_{E}$W and QPG$\Gamma_{E}$W.
The summations over the frequencies $\omega$ and $\nu$ in
(\ref{eq40}) were performed for $|\omega|<20$ and $0\leq\nu<20$.

\begin{figure}[b]
\centering
\includegraphics[width=9.0 cm]{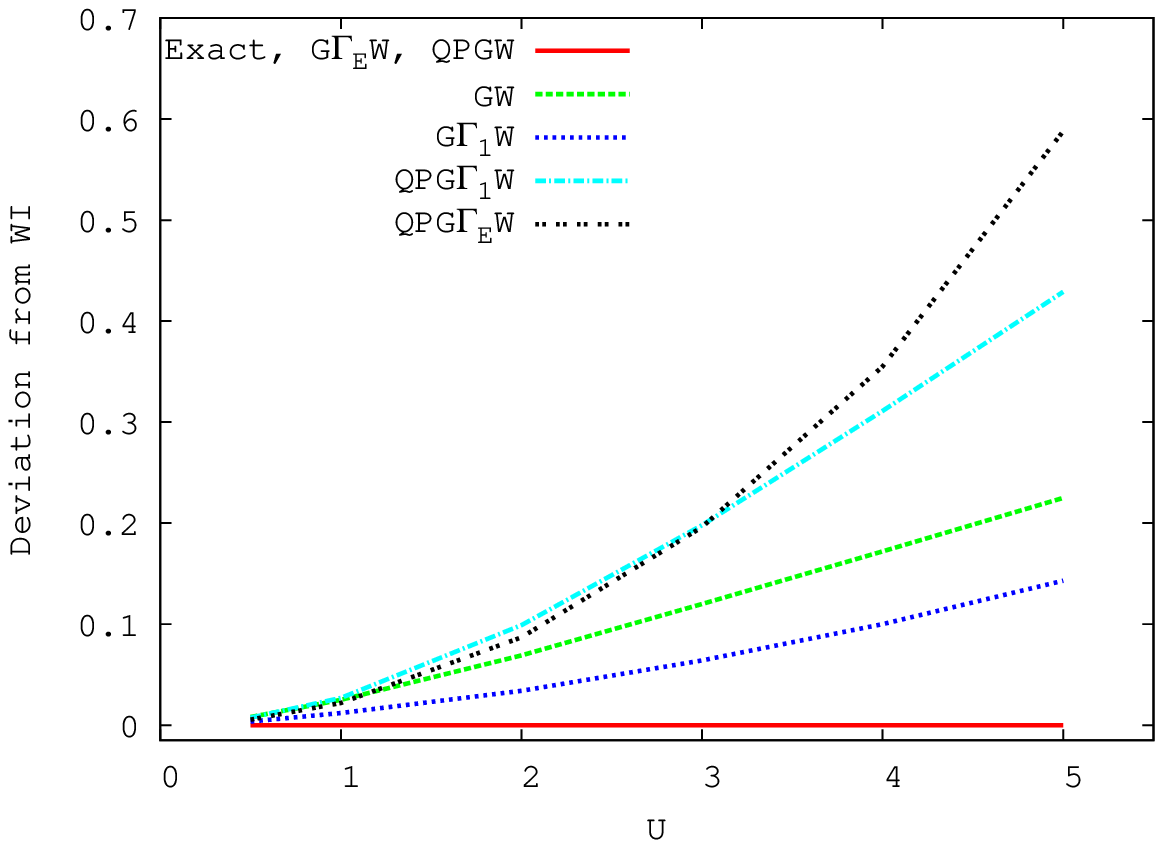}
\caption{(Color online) $\mathbf{q}\rightarrow 0$ Ward Identity
average violation as a function of U parameter.} \label{q0_WI_2}
\end{figure}

\begin{figure}[t]
\centering
\includegraphics[width=9.0 cm]{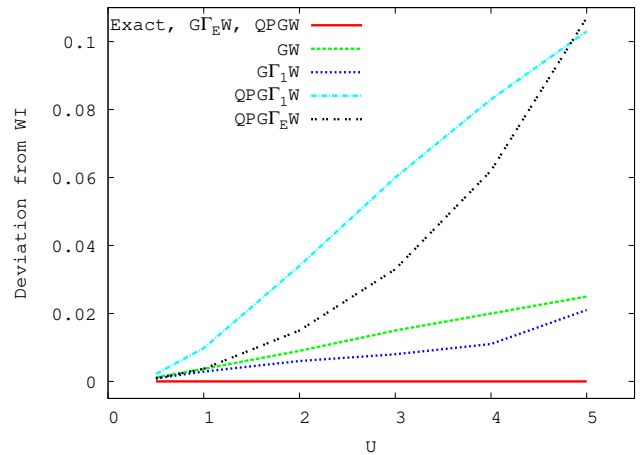}
\caption{(Color online) $\mathbf{q}\rightarrow 0, \nu\rightarrow 0$
Ward Identity average violation as a function of U parameter.}
\label{q0nu0_WI_2}
\end{figure}

The figure \ref{q0_WI_2} shows the similar plot obtained with the
long-wave limit of the WI (\ref{eq31d},\ref{eq31e}). The formula
analogous to (\ref{eq40}) has been used but without the
$k$-summation. Finally in the figure \ref{q0nu0_WI_2} the deviation
from the long-wave+static limit of the WI (\ref{eq31f},\ref{eq31g})
is shown. In this case, the summation over $\nu$ has not been
included.

One can do a few observations from the results presented.

Most important for the present work is an observation following from
the errors in Green's function, that the QPG$\Gamma_{E}$W method
doesn't show any noticeable improvement as compared to the other
approximate methods studied. Opposite to that - with U increasing it
quickly becomes the worst of the methods considered (especially for
the higher frequencies). The observation perfectly correlates with
the results obtained for the internal energy and for the Ward
Identities (especially with the results for the full WI). It means
that putting the exact vertex together with the QP self-consistency
is not compatible.

Grouping the methods based on the QP-approximation in one group and
the rest of the methods in another (they satisfy the DE), one can do
another conclusion. Namely for the frequencies larger than
approximately 1.5 the QP group has considerably larger errors in G
than the group with the full self-consistency. This again correlates
strongly with the internal energy graph and with the full WI.

Further, in the methods satisfying the Dyson equation the increasing
in the accuracy of the vertex
($1\rightarrow\Gamma_{1}\rightarrow\Gamma_{E}$) consistently
improves the accuracy of G, finally making it exact when the exact
vertex is being applied. Naturally, the same tendency can be noticed
looking at the internal energy graph and at all WI graphs.

Opposite to that, in the QP-based methods the situation is more
complicated. At lowest U (0.5 and 1) the QPGW is the best and the
QPG$\Gamma_{E}$W is the worst at low frequencies with the reversed
tendency at high frequencies. For U equal 2, 3, and 4 one can notice
that now  the QPG$\Gamma_{E}$W is the best and the QPGW is the worst
at low frequencies with the reversed tendency at high frequencies.
For U equal 5 the accuracy of the QPG$\Gamma_{E}$W approach
deteriorates also at low frequencies and its failing at higher
frequencies becomes severe. The important point here is, that the
above high-frequency tendencies in the QP-based methods correlate
well with the tendencies in the internal energy as one can easily
see. Namely, for U$\leq$1 the the QPG$\Gamma_{E}$W gives the best
(among the QP-based schemes) internal energy and for U$>$1 it is the
worst among them. This also correlates very well with the deviations
from the full WI.

For U=0.5 the QPGW approach produces the best G at lowest
frequencies among all approximate methods. One can speculate, that
this fact is similar to the well known fact, that in weakly
correlated real materials the QPGW is probably the best approach for
studying the spectra (which are defined by the low frequency
behavior of Green's function). This fact can be related with the
exact fulfillment of the long-wave limit of the WI in the QPGW
approach.

An interesting observation comes from the figure for U=1. As one can
see there is a strict separation in the accuracy of G obtained
within the QP-group and within the non QP-group. The reasons for
this feature have not been studied in this work but it could be
related to the above mentioned swapping between different tendencies
in the QP-based methods at this U.

The GW-based (without the QP approximation) methods give numerically
exact internal energy till $U\approx2$ which is not very surprising
because they are conserving (in Baym-Kadanoff
sense\cite{pr_124_287}) and in the weakly correlated regime should
produce accurate total energies. Generally it is not a good idea to
use the QP-based methods to evaluate the total energy because even
if they originate from the same $\Psi$-functional\cite{ijmpb_13_535}
as the GW-based methods  the Dyson equation (DE) is not satisfied in
them anymore. As it was indicated above, the deviations from the
exact Green's function in the QP-based approaches are especially
noticeable at high frequencies which are important for the total
energies evaluation. The results shown in Fig.\ref{E_2} clearly
support this point of view.

Internal energy obtained in QP-based approximations deviates
significantly from the exact one even at small U. It is interesting
that in the QPGW scheme the error remains almost unchanged till
$U=5$ which probably is accidental because at $U\geq3$ the PT
becomes unreliable. The important difference between the GW-based
and the QPGW-based methods is however that the G$\Gamma_{1}$W
improves the internal energy as compared to the GW, whereas applying
the vertex of improved accuracy
($1\rightarrow\Gamma_{1}\rightarrow\Gamma_{E}$) in the sequence
QPGW$\rightarrow$QPG$\Gamma_{1}$W$\rightarrow$QPG$\Gamma_{E}$W makes
the results worse and worse which also tells us that one should not
apply the vertex correction combined with the QP approximation.

As it is clear, the degree of violation of the WI correlates well
with the errors in the Green functions at average and high
frequencies, which are responsible for the accuracy of the
calculated total energies. And indeed, the comparison of the Figures
\ref{E_2} and \ref{F_WI_2} tells us, that the accuracy of the total
energy and the accuracy of the full WI fulfillment are closely
related. Thus, the full WI can be useful as a measure of the
accuracy of the calculated total energies.

There was a hope, that the deviation from the long-wave+static limit
of the WI correlates well with the errors in G at low frequencies.
But for the two-site Hubbard model this seems to be not the case
(besides the above mentioned success of the QPGW approach at the
smallest values of U). However, in real materials where the spectra
obtained with the QPGW are generally noticeably better than the
spectra obtained with the scGW the situation might be different.

\section{Conclusions} \label{concl}

In this work the Hedin's equations\cite{pr_139_A796} for the
two-site Hubbard model have been solved self-consistently with and
without applying the quasiparticle approximation for the Green
function. The study has been performed both when the exact
three-point vertex function was used as an input and when the
perturbative theory (in its zero and first orders in W) was used to
evaluate the corresponding vertices self-consistently. The results
of this work obtained with the exact vertex have direct impact on
what one can expect from the future implementation of the sc GW+DMFT
and the sc QPGW+DMFT schemes. As it has been shown here, only the
GW+DMFT approach can be considered as useful approximation. However,
as it was said in the Introduction, this work deals with an ideal
situation, when the GW part and the DMFT part are perfectly
separable and the subspace where GW is used is very weakly
correlated (so that GW and QPGW give identical results for the
weakly correlated subspace). In practice, it is not always the case.
The correlations in the "GW"-subspace might be noticeable. In such
situation, the QPGW might be superior (with respect to the GW) for
the subspace not included in the DMFT part. The conclusions about
the DMFT part obviously remain as before - one should not impose the
QP approximation on G in the DMFT part. As it seems, in such
circumstances the preference should be given to the approach
(GW-based or QPGW-based) depending on which subspace (the "weakly"
or the "strongly" correlated) is more important for the problem
under consideration. However, on the fundamental level, such a
situation should be resolved either by the increasing of the
subspace covered by the DMFT part or (which seems to be easier
practically) by including more diagrams beyond GW for the "weakly"
correlated subspace.

It has been shown, that the methods with the PT-based vertices (when
they are applicable) reveal similar tendencies (for example if one
chooses between the QP self-consistency and the full sc) as the
methods based on the exact vertices. Namely, when the correlation
strength increases, both the PT-based and the exact vertices-based
schemes begin to fail if one uses the QP self-consistency. Also of
practical importance is the finding that the violation of the WI in
any particular method correlates well with the general applicability
of the given method. This can be an useful information if one sees
to apply the schemes with vertex corrections to the real materials
where the exact solutions are not available to serve as a judgement.

The results of this work are of the methodological importance. Of
course the two-site HM doesn't cover all possible regimes of
correlations which may happen in realistic materials. In order to
cover a little bit more of the possible regimes of correlations the
similar work on the homogeneous electron gas is now being performed
(with the three-point vertex functions calculated within the PT
only).


\begin{thebibliography}{14}
\expandafter\ifx\csname
natexlab\endcsname\relax\def\natexlab#1{#1}\fi
\expandafter\ifx\csname bibnamefont\endcsname\relax
  \def\bibnamefont#1{#1}\fi
\expandafter\ifx\csname bibfnamefont\endcsname\relax
  \def\bibfnamefont#1{#1}\fi
\expandafter\ifx\csname citenamefont\endcsname\relax
  \def\citenamefont#1{#1}\fi
\expandafter\ifx\csname url\endcsname\relax
  \def\url#1{\texttt{#1}}\fi
\expandafter\ifx\csname urlprefix\endcsname\relax\def\urlprefix{URL
}\fi \providecommand{\bibinfo}[2]{#2}
\providecommand{\eprint}[2][]{\url{#2}}

\bibitem[{\citenamefont{{P.~Sun and G.~Kotliar}}(2002)}]{prb_66_085120}
\bibinfo{author}{\bibnamefont{{P.~Sun and G.~Kotliar}}},
  \bibinfo{journal}{Phys.~Rev.B} \textbf{\bibinfo{volume}{66}},
  \bibinfo{pages}{085120} (\bibinfo{year}{2002}).

\bibitem[{\citenamefont{{S.~Biermann, F.~Aryasetiawan, and
  A.~Georges}}(2003)}]{prl_90_086402}
\bibinfo{author}{\bibnamefont{{S.~Biermann, F.~Aryasetiawan, and A.~Georges}}},
  \bibinfo{journal}{Phys.~Rev.~Lett.} \textbf{\bibinfo{volume}{90}},
  \bibinfo{pages}{086402} (\bibinfo{year}{2003}).

\bibitem[{\citenamefont{{J.~M.~Tomczak, M.~Casula, T.~Miyake, and
  S.~Biermann}}(2014)}]{prb_90_165138}
\bibinfo{author}{\bibnamefont{{J.~M.~Tomczak, M.~Casula, T.~Miyake, and
  S.~Biermann}}}, \bibinfo{journal}{Phys.~Rev.~B}
  \textbf{\bibinfo{volume}{90}}, \bibinfo{pages}{165138}
  (\bibinfo{year}{2014}).

\bibitem[{\citenamefont{{A.~N.~Rubtsov, V.~V.~Savkin, and
  A.~I.~Lichtenstein}}(2005)}]{prb_72_035122}
\bibinfo{author}{\bibnamefont{{A.~N.~Rubtsov, V.~V.~Savkin, and
  A.~I.~Lichtenstein}}}, \bibinfo{journal}{Phys.~Rev.~B}
  \textbf{\bibinfo{volume}{72}}, \bibinfo{pages}{035122}
  (\bibinfo{year}{2005}).

\bibitem[{\citenamefont{{P.~Werner, A.~Comanac, L.~de~Medici, M.~Troyer, and
  A.~J.~Millis}}(2006)}]{prl_97_076405}
\bibinfo{author}{\bibnamefont{{P.~Werner, A.~Comanac, L.~de~Medici, M.~Troyer,
  and A.~J.~Millis}}}, \bibinfo{journal}{Phys.~Rev.~Lett.}
  \textbf{\bibinfo{volume}{97}}, \bibinfo{pages}{076405}
  (\bibinfo{year}{2006}).

\bibitem[{\citenamefont{{A.~Kutepov, K.~Haule, S.~Y.~Savrasov, and
  G.~Kotliar}}(2012)}]{prb_85_155129}
\bibinfo{author}{\bibnamefont{{A.~Kutepov, K.~Haule, S.~Y.~Savrasov, and
  G.~Kotliar}}}, \bibinfo{journal}{Phys.~Rev.~B} \textbf{\bibinfo{volume}{85}},
  \bibinfo{pages}{155129} (\bibinfo{year}{2012}).

\bibitem[{\citenamefont{{A.~Kutepov, S.~Y.~Savrasov, and
  G.~Kotliar}}(2009)}]{prb_80_041103}
\bibinfo{author}{\bibnamefont{{A.~Kutepov, S.~Y.~Savrasov, and G.~Kotliar}}},
  \bibinfo{journal}{Phys.~Rev.~B} \textbf{\bibinfo{volume}{80}},
  \bibinfo{pages}{041103} (\bibinfo{year}{2009}).

\bibitem[{\citenamefont{{T.~Kotani and M.~van~Schilfgaarde,
  S.~V.~Faleev}}(2007)}]{prb_76_165106}
\bibinfo{author}{\bibnamefont{{T.~Kotani and M.~van~Schilfgaarde,
  S.~V.~Faleev}}}, \bibinfo{journal}{Phys.~Rev.B}
  \textbf{\bibinfo{volume}{76}}, \bibinfo{pages}{165106}
  (\bibinfo{year}{2007}).

\bibitem[{\citenamefont{{M.~van~Schilfgaarde, T.~Kotani, and
  S.~Faleev}}(2006)}]{prl_96_226402}
\bibinfo{author}{\bibnamefont{{M.~van~Schilfgaarde, T.~Kotani, and
  S.~Faleev}}}, \bibinfo{journal}{Phys.~Rev.~Lett.}
  \textbf{\bibinfo{volume}{96}}, \bibinfo{pages}{226402}
  (\bibinfo{year}{2006}).

\bibitem[{\citenamefont{{J.~M.~Tomczak, M.~van~Schilfgaarde, and
  G.~Kotliar}}(2012)}]{prl_109_237110}
\bibinfo{author}{\bibnamefont{{J.~M.~Tomczak, M.~van~Schilfgaarde, and
  G.~Kotliar}}}, \bibinfo{journal}{Phys.~Rev.~Lett.}
  \textbf{\bibinfo{volume}{109}}, \bibinfo{pages}{237110}
  (\bibinfo{year}{2012}).

\bibitem[{\citenamefont{{J.~Tomczak}}(2014)}]{arx.1411.5180}
\bibinfo{author}{\bibnamefont{{J.~Tomczak}}},
  \bibinfo{journal}{arXiv.cond.mat.:1411.5180}  (\bibinfo{year}{2014}).

\bibitem[{\citenamefont{{L.~Hedin}}(1965)}]{pr_139_A796}
\bibinfo{author}{\bibnamefont{{L.~Hedin}}}, \bibinfo{journal}{Phys.~Rev.}
  \textbf{\bibinfo{volume}{139}}, \bibinfo{pages}{A796} (\bibinfo{year}{1965}).

\bibitem[{\citenamefont{{G.~Baym, and L.~P.~Kadanoff}}(1961)}]{pr_124_287}
\bibinfo{author}{\bibnamefont{{G.~Baym, and L.~P.~Kadanoff}}},
  \bibinfo{journal}{Phys.~Rev.} \textbf{\bibinfo{volume}{124}},
  \bibinfo{pages}{287} (\bibinfo{year}{1961}).

\bibitem[{\citenamefont{{C.-O.~Almbladh, U.~von~Barth and
  R.~van~Leeuwen}}(1999)}]{ijmpb_13_535}
\bibinfo{author}{\bibnamefont{{C.-O.~Almbladh, U.~von~Barth and
  R.~van~Leeuwen}}}, \bibinfo{journal}{Int.~J. of Mod.Phys. B}
  \textbf{\bibinfo{volume}{13}}, \bibinfo{pages}{535} (\bibinfo{year}{1999}).

\end{thebibliography}

\end{document}